\documentclass[12pt]{iopart}
\usepackage{graphicx}
\begin{document}

\title[Critical point density]{Density of critical points for a Gaussian random function}
\author{H. Vogel and W. M\"ohring}
\address{ Max Planck Institute for Dynamics and Self-Organization, \\
Bunsenstr. 10, 37073 G\"ottingen, Germany}
\ead{wmoehri@gwdg.de}

\begin{abstract}
 Critical points of a scalar quantitiy are either extremal points or saddle points. The character of the critical points is determined by the sign distribution of the eigenvalues of the Hessian matrix. For a two-dimensional homogeneous and isotropic random function topological arguments are sufficient to show that all possible sign combinations are equidistributed or with other words, the density of the saddle points and extrema agree. This argument breaks down in three dimensions. All ratios of the densities of saddle points and extrema larger than one are possible. For a homogeneous Gaussian random field one finds no longer an equidistribution of signs, saddle points are slightly more frequent.
\end{abstract}
%\submitto{\JPA}
%\maketitle
\section{Introduction}

Often turbulent phenomena are accompanied by random fluctuations of some scalar quantity. One may think of the density, temperature or optical
refraction index for compressible flows, concentrations in the case of mixtures
or reaction or combustion rates in the case of chemically active substances.
Important properties of these fields are their critical points, i.e. those
points where their gradient vanishes. They give rise to a subdivision of these fields into "dissipation elements" \cite{l0} very similar to the partition of the velocity field into "eddies" generated from the stagnation points \cite{l00,l01}. One should also mention \cite{l9} where the routes to the generation of critical points out of simpler flows are studied. 
In addition to their importance for the characterization of the field, they are often also of direct physical
significance. Let us discuss the situation in more detail in case of a temperature field. 
There the heat flux is directed towards lower temperatures, i.e. the heat flux lines end in a temperature minimum. Each temperature minimum is surrounded by a cell consisting of points with heat flux directed towards the minimum. It does of course not mean that the actual heat transport in the turbulent flow is appropriately described by this cell structure as the cells change from time to time and there usually is a significant contribution from convection to the heat transport. The cell structure is more suited to visualize the temperature field. As the cells are associated with temperature minima the distribution of the minima is of special interest. The distribution of the minima depends of course on the random properties of the turbulence and is therefore difficult to obtain. Some hints can however be obtained if one assumes Gaussian random fields, an assumption which is often made in turbulence studies. The properties of the critical points depend on the Hessian matrix of the scalar field. The Hessian of isotropic homogeneous random functions has been studied for two-dimensional fields by Longuet-Higgins \cite{N1} in a paper on light reflection at a random surface and for three-dimensional fields by Halperin and Lax \cite{N2} in a paper on impurities in semiconductors. In these early works the method of Rice \cite{N3} to represent a random function as a superposition of trigonometric functions is used. This method assumes implicitly homogeneity of the random function, its role was therefore not considered explicitly in these works. More recently the Hessians of large dimensions were studied in the context of random matrix theory \cite{l1} by Fyodorov \cite{N4,N4a}. The relation of the Hessians to the Gaussian orthogonal ensemble (GOE) was clearly formulated in this work and it was shown that one can introduce a normally distributed auxiliary variable $t$ such that the ensemble of Hessians of homogeneous isotropic random functions is obtained from GOE by averaging over this variable. Through this approach it is possible to apply results from GOE studies at the cost of one extra average (i.e. integration). Fyodorov shows how this can effectively be done. 

The topic which is of high interest to us is the distribution of extrema in the critical points.
The type of the critical point is determined by the sign distribution of the eigenvalues of the Hessian. Usually the number of negative eigenvalues is called the index of the matrix. Methods of functional integration have been used by Bray and Dean \cite{N5} to determine the asymptotic distribution of the index and Fyodorov , Sommers and Williams \cite{N6} have determined the asymptotic value of the minima i.e. the critical points of index zero. Comparing their eqs. (31) and (23) one finds that the probability of a critical point to be a minimum is asymptotically given by exp(-N) for N-dimensional homogeneous random functions.

The situation is completelely different for two-dimensional random functions. There it is well known that it follows from topological reasons or from the vanishing of the average of the determinant of the Hessian that the densities of the extrema and of the saddle points agree. This is e.g. the basis of the Poincare-index and of the Brouwer-degree.\cite{l4,l5} For random functions with even probability distribution, e.g. in the  Gaussian case one has one half of the critical points with index 1 and a quarter with index zero or two. The four different sign combinations of the eigenvalues are equidistributed. On the other hand the result of Fyodorov et al shows that the fraction of critical points which are minima, namely exp(-N), is for large matrices well below the fraction of $2^{-N}$ to be expected for an equidistribution of signs.

Our interest is the three-dimensional case. There the strong topological restrictions from the two-dimensional case no longer apply. From Morse theory \cite{l5a} one can conclude that the density of the extrema cannot be smaller than the number of saddle points, but equality, as in two dimensions is no longer true. In an appendix, we describe functions having various fractions of minima. So we consider Gaussian random functions. Although we could use results of Fyodorov for the critical point density, the situation is more complicated for the density of minima. So we stay with the ensemble of Hessians and use the methods developped in random matrix theory as far as they apply and make use of the additional symmetry which follows from the homogeneity of the random functions. We think that this symmetry can best be expressed in terms of the wedge product of the exterior algebra. One finds that the average of all wedge products of column vectors of the Hessian vanish. For the three-dimensional case one can then evaluate the remaining integrals by fairly elementary means and obtains for the fraction of minima under all critical points the value $1/4-3\sqrt6/58= 0.1233$ slightly less than $1/8$ to be expected for equidistribution of signs. 

 \section{Basic relations}

To obtain the density of the critical points of a scalar random function $C(\bi{x})$, i.e. points in  $N$-dimensional $\bi{x}$-space where the gradient $\bi{g}=\nabla C$ vanishes one observes that
each critical point gives a contribution $1/|\det({\bf H}(\bi{x}))|$  to the integral over $\delta(\bi{g})$, where $\delta(\bi{g})$ denotes the Dirac $\delta$-function. Here ${\bf H}(\bi{x})$ denotes the Jacobian matrix of $\bi{g}$, i.e. the Hessian matrix of the second derivatives of $C$ and $\det({\bf H})$ its determinant.
Then the density of the critical points $n_{cr}$ can be obtained as
\begin{equation} n_{\rm cr}=<|\det({\bf H})|\,\delta(\bi{g})>\qquad {\rm with} \quad \bi{g}=\bigg(\frac{\partial C}{\partial x_i}\bigg), \, {\bf H}=\bigg(\frac{\partial^2 C}{\partial x_i\partial x_k}\bigg),
\label{g1} 
\end{equation}
where $<a>$ denotes the average of $a$. This is the so-called Kac-Rice formula. Similarly the density $n_{\rm mi}$ of the minima of $C$ can be obtained from
\begin{equation} n_{\rm mi}=<\chi_{\rm mi}({\bf H})|\det({\bf H})|\delta(\bi{g})>.\label{g2} 
\end{equation}
Here $\chi_{\rm mi}({\bf H})$ denotes the characteristic function of the minima which is one if all eigenvalues of ${\bf H}$ are positive and otherwise zero.

From the homogeneity condition one obtains with integration by parts
\begin{equation}
< H_{ii}H_{kk}> = <H_{ik}^2>. \label{g3} 
\end{equation} 
This is in marked contrast random matrix theory where matrices ${\bf K}$ from the GOE fufill
\begin{equation}
< K_{ii}K_{kk}>_{GOE} = 2\delta_{ik}<K_{ik}^2>_{GOE}. \label{g3a} 
\end{equation} 
The isotropy of $C(\bi{x})$ lead with \eref{g3} to the probability distribution function (pdf) \cite{N4}
\begin{eqnarray}
P & = & c\exp\Big(-\alpha \bi{g}^2 -\beta \big(\Tr {\bf H}^2-\frac{1}{N+2}(\Tr{\bf H})^2\big)\Big) \label{g4} \\
& &  {\rm with} \quad \alpha=\frac1{2<g_1^2>}, \; \beta=\frac1{2<H_{12}^2>}. 
\nonumber 
\end{eqnarray}
where $\Tr$ denotes the trace of the matrix and $c$ is a normalization constant. Obviously all off-diagonal elements are independant and have the same normal distribution, exactly as for GOE. Differences occur only in the diagonal elements. Therefore Fyodorov introduced an additional normally distributed variable $t$ (averages over $t$ are denoted by a subscript NO) and wrote with ${\bf K}$ from GOE
\begin{equation}
H_{ik}=K_{ik}+t\delta_{ik}. \label{g4a} 
\end{equation}
To determine the mean square of $t$ one averages over GOE and NO to obtain for $i\neq k$
\begin{equation}
<H_{ik}^2>=<H_{ii}H_{kk}>=<(K_{ii}+t)(K_{kk}+t)>_{GOE,NO}=<t^2>_{NO}
\end{equation}
Alternatively it is possible to assume $t=\chi \Tr {\bf K}$ and one obtains for $i\neq k$
\begin{eqnarray*}
 && <H_{ii}H_{kk}>=<(K_{ii}+\chi\Tr{\bf K})(K_{kk}+\chi\Tr{\bf K})>_{GOE} = \\ 
 && 2\chi<K_{ii}^2>_{GOE}+\chi^2 N <K_{ii}^2>_{GOE}=<H_{ik}>^2=<K_{ik}^2>_{GOE}
\end{eqnarray*}
and therefore
\begin{displaymath}
2N\chi^2+4\chi -1 = 0.
\end{displaymath}
This representation is useful for Monte Carlo calculations.

Now we want to show that the homogeneity condition \eref{g3}
implies that the average value of the determinant of ${\bf H}$ vanishes. Much more is true, namely the average of all coefficients of the characteristic polynomial of ${\bf H}$ vanish (with the obvious exception of the first one which is equal to one). This is in marked contrast to the GOE where already the average of the determinant is different from zero. This property of ${\bf H}$ can be conveniently shown using  the exterior product of the exterior algebra \cite{l2}. This is a product $\bi{h_1}\wedge \bi{h_2}$ defined  for $p=1,2,\dots , N$ on the vector space of $p$-tuples $\bi{v_1} = \bi{h_1 \wedge h_2 \wedge \dots \wedge h_p}$ of vectors from some basic vector space of vectors $\bi{h_i}$ which fulfills the associative and distributive laws and furthermore an anticommutative law
\[
\bi{h_1} \wedge \bi{h_2} = - \bi{h_2} \wedge \bi{h_1}.
\]
As basic vector space we take an $N$-dimensional vector space with basic vectors $\bi{e_1}.\bi{e_2},\dots ,\bi{e_N}$ and consider the row vectors of ${\bf H}$, i.e. $\bi{h_i}= H_{i1}\bi{e_1}+H_{i2}\bi{e_2}+\dots + H_{iN}\bi{e_N}$. Then we can write the homogeneity condition \eref{g3} as
\begin{eqnarray*}
<\bi{h_i} \wedge \bi{h_k}> & = & <(H_{ii}\bi{e_i}+H_{ik}\bi{e_k}) \wedge (H_{ki}
\bi{e_i}+H_{kk}\bi{e_k})> \nonumber \\ & = & <H_{ii}H_{kk}-H_{ik}H_{ki}>\bi{e_i} \wedge \bi{e_k} =0,
\end{eqnarray*}
i.e. the wedge product of two arbitrary row vectors of ${\bf H}$ has vanishing mean. As the average of an arbitrary number $p$ of Gaussian random variables vanishes for odd $p$ and is for even $p$ just the sum of all partitions of the product into pairs and taking their averages one finds that all these averages vanish too,
\begin{equation}
<\bi{h_{i_1}} \wedge \bi{h_{i_2}} \wedge \dots \wedge \bi{h_{i_p}}> =0 \qquad p=1,2,\dots , N. \label{g5a} 
\end{equation}
These equations show the high degree of symmetry of this Gaussian matrix ensemble, all antisymmetric products have vanishing mean. One special case, namely $p=N$, states the vanishing of the average determinant,
\begin{equation}
<\det({\bf H})>\; =0.\label{g5b} 
\end{equation} 
From \eref{g5a} one can obtain a much more general equation. With \eref{g5a} one has also
\[\fl
<(\bi{h_{i_1}}-\bi{a_{i_1}}) \wedge (\bi{h_{i_2}}-\bi{a_{i_2}}) \wedge \dots \wedge (\bi{h_{i_N}}-\bi{a_{i_N}})>\; =(-1)^N
\bi{a_{i_1}} \wedge \bi{a_{i_2}} \wedge \dots \wedge \bi{a_{i_N}}
\]
with arbitrary constant vectors $\bi{a}_p$. This can also be written as
\begin{equation}
<\det({\bf H}-{\bf A})>\;=\det({\bf -A})=(-1)^N\det({\bf A}).\label{g5c} 
\end{equation} 
with an arbitrary constant matrix ${\bf A}$. For ${\bf A}$ proportional to the identity matrix
\eref{g5c} shows, that the average of all nontrivial cofficients of the characteristic polynomial of ${\bf H}$ vanishes. Related results for GOE can be found in \cite{l1,l3} where the moments of the characterstic polynomial have been determined. With \eref{g4a}
an alternative derivation is possible.

An interesting application of \eref{g5b} is possible in two dimensions. There the sign of $\det({\bf H})$ is positive for extrema of $C$ and negative for saddle points. Then we see from \eref{g2} that \eref{g5b} implies equal densities of the extrema and of the saddle points in two dimensions, this argument, however, cannot be carried over to higher dimensions, there one has equal densities of critical points of even and odd indices (Brouwer degree).

\section{Three-dimensional matrices}

From \eref{g4} one notices that $\bi{g}$ and ${\bf H}$ are statistically independent. So the density of the critical points and of the maxima can be written as a product of two factors $n=n^{(1)}\,n^{(2)}$, where the first factor depends only on the average values of $\bi{g}$ and the second of those of ${\bf H}$. The first factor is easily determined
\begin{equation}
n^{(1)}=\frac{\int \delta(\bi{g})\exp(-\alpha\bi{g}^2)\,\rmd^3\bi{g}}
{\int \exp(-\alpha\bi{g}^2)\,\rmd^3\bi{g}} =\sqrt{\frac{\alpha}{\pi}}^{\displaystyle3} \label{g5} 
\end{equation} 
It remains to find the second factor.

Let us now restrict ourselves to three dimensions, $N=3$. The pdf for the Hessian is then given by
\begin{equation}
P=c\exp(-\beta (\Tr {\bf H}^2- (\Tr{\bf H})^2/5)) \qquad {\rm with}\quad
\beta=\frac1{2<H_{12}^2>}.\label{g6} 
\end{equation} 

We are interested in averages depending only on the eigenvalues of ${\bf H}$. Then it is common \cite{l1} to introduce the eigenvalues $E_1,E_2,E_3$ as new variables and to average over the remaining variables. As the ordering of the eigenvalues is arbitrary, we assume 
\begin{equation}
E_1\leq E_2 \leq E_3. \label{g7} 
\end{equation} 
Then it is shown in \cite{l1} that the pdf in the eigenvalue space $P_e$ is given by
\[
P_e=c\exp(-\beta\bi{e}^T {\bf A}\bi{e})\Delta(\bi{e})\quad {\rm with} \;
A_{ik}=\frac15( 5\delta_{ik}-1),
\] 
where $\Delta(\bi{e})=(E_3-E_2)(E_2-E_1)(E_3-E_1)$ and $\bi{e}$ is a column vector of the eigenvalues.

Now we introduce new variables $y_1,y_2,y_3$ through
\begin{equation}
\bi{e}={\bf B}\left(\begin{array}{c}y1\\y2\\y_3
\end{array}\right) =
\frac1{\sqrt{6}}\left(\begin{array}{ccc}-2 & 0 & \sqrt5\\1 & -\sqrt3 & \sqrt5 \\
1 & \sqrt3 & \sqrt5
\end{array}\right)\bi{y}. \label{g8}
\end{equation}
One verifies easily that ${\bf B}^T{\bf A B}$ is the identity matrix. Therefore the pdf reads in the new variables
\[
P_e=c\exp(-\beta\bi{y}^T \bi{y})\Delta({\bf B}\bi{y}).
\]
With spherical polar coordinates
\[y_1=r\cos\varphi\sin\vartheta,\, y_2=r\sin\varphi\sin\vartheta,\, y_3=r\cos\vartheta \]
one obtains $\Delta(\bi{e})=r^3\Delta_3(\vartheta,\varphi)=r^3\sin^3\vartheta\sin(3\varphi)/
\sqrt{2}$.
As we are interested in the average of homogeneous quantities $\Phi(\bi{e}) = r^n\Phi_n(\vartheta,\varphi)$ only, we can write
\[ <\Phi>\; = \frac{c}{\sqrt{2}} \int r^{n+5}\exp(-\beta r^2)\Phi_n(\vartheta,\varphi) \sin(3\varphi) \, \sin^4\vartheta \;\rmd r \rmd\varphi \rmd\vartheta. \]
Now we can perform the $r$-integration and obtain
\begin{equation}
<\Phi>\; = \frac{c\Gamma(3+n/2)}{\sqrt{2}\beta^{3+n/2}}\int \Phi_n(\vartheta,\varphi) \sin(3\varphi) \, \sin^4\vartheta \; \rmd\varphi \rmd\vartheta,\label{g9} 
\end{equation} 
where $\Gamma$ denotes the $\Gamma$-function.

Now we have reduced the integral from one in the 6-dimensional space of the $H_{ik}$ to an integral over a 2-dimensional sphere. One finds easily, that $E_1=E_2$ corresponds to $\varphi=\frac{\pi}3$ and $E_2=E_3$ to $\varphi=0$. Then, because of the restriction \eref{g7} the integration is not over the whole sphere but only over the segment 
\begin{equation}
0\leq \varphi \leq \pi/3,\qquad  0\leq \vartheta \leq \pi.\label{g10} 
\end{equation}
\begin{figure}
	\begin{center}
	\includegraphics[scale=0.6,bb=0 0 319 320]{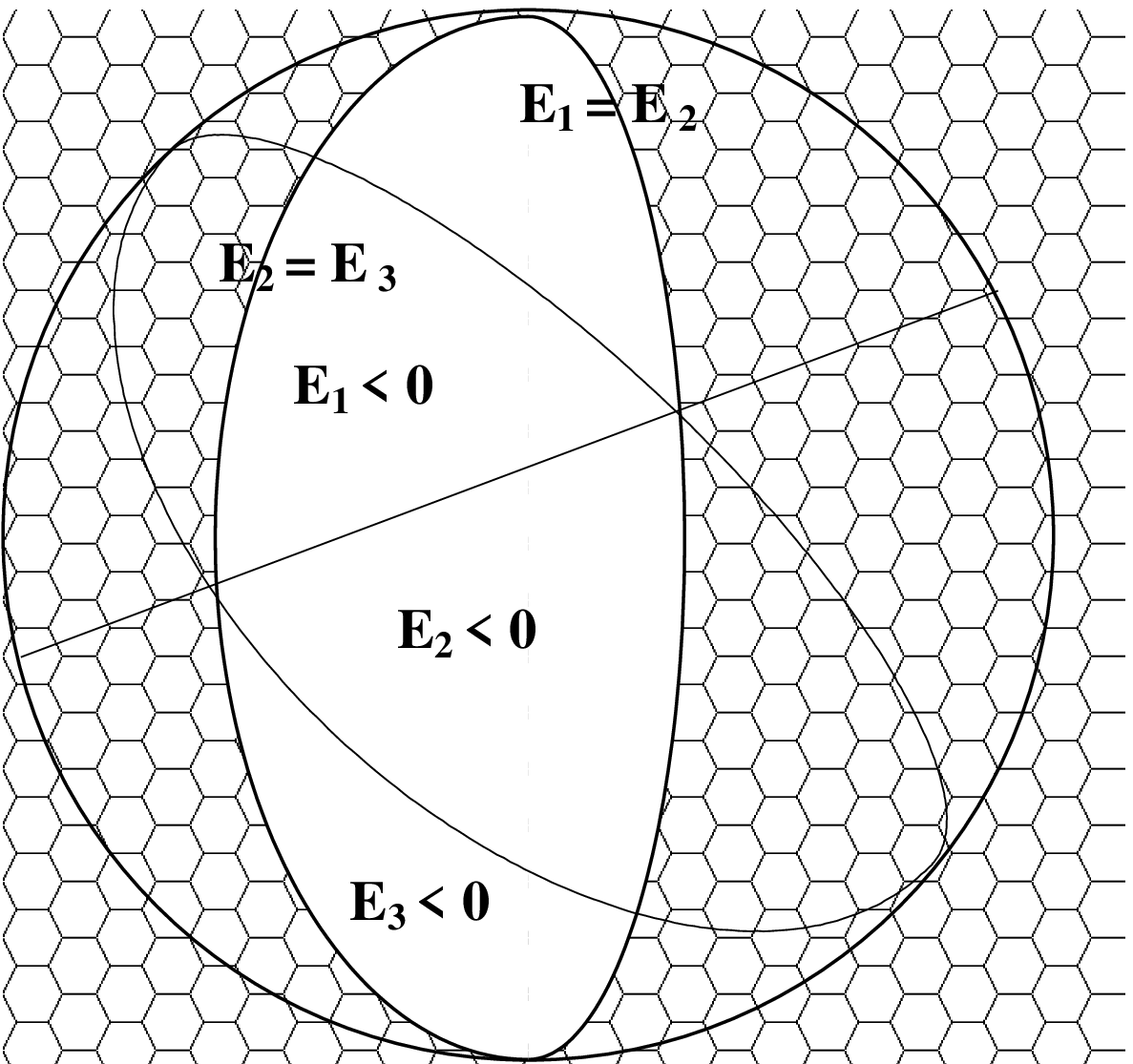}
	\end{center}
\caption{ The unit sphere in $y$-space. Bright is the physically significant region \eref{g7}. Shown is also the partition into regions of only positive eigenvalues (topmost region), one negative eigenvalue ($E_1<0$), two negative eigenvalues ($E_2<0$) and all eigenvalues negative.}
\end{figure}
The other regions of the sphere correspond to a different ordering of the eigenvalues.As the enumeration of the eigenvalues is usually arbitrary, one can as well integrate over the whole sphere and divide the result by the number of different orderings, namely six. The configuration is shown in figure 1 where we additionally show the cross section of the sphere with the planes $E_1=0$, $E_2=0$, and $E_3=0$. As these planes, as well as $E_1=E_2$ and $E_3=E_2$, all pass  through the origin, the cross sections are all great circles. The segment \eref{g7} is then divided into four spherical triangles. The topmost is bounded in addition to \eref{g10} by $E_1=0$, it corresponds therefore to eigenvalues which are all positive, the next triangle is bounded by $E_1=0$ and $E_2=0$, corresponds therefore to one negative eigenvalue, the next one to two negative eigenvalues and the last one to eigenvalues which are all negative. One also notices that the segment and the pdf is not changed if the signs of the eigenvalues and their numeration is reversed. The four spherical triangles are however interchanged. Therefore the first and the fourth and the second and the third lead often to contributions of the same size.

\section{Critical points}

As a first step, we determine the normalization constant $c$ of the pdf. As usual, it can be determined from the average of one which is one. As one is a homogeneous function of degree $n=0$ one has from \eref{g9}
\begin{equation}
1= \frac{c\Gamma(3)}{\sqrt{2}\beta^{3}}\int_0^{\pi/3} \sin(3\varphi)\,\rmd\varphi \,\int_0^{\pi} \sin^4\vartheta\, \rmd\vartheta=\frac{c\pi}{2\sqrt{2}\beta^3}.  \label{g11} 
\end{equation} 
Now we determine the second factor $n^{(2)}$ for the spatial density of the minima. As the determinant of ${\bf H}$ is a homogeneous function of third degree we obtain from \eref{g9}
\begin{equation}
n^{(2)}_{\rm mi}=<\chi_{\rm mi}({\bf H})\det({\bf H})> = \frac{c\Gamma(9/2)}{\sqrt{2}\beta^{9/2}} \int \det({\bf H}) \sin(3\varphi) \, \sin^4\vartheta \; \rmd\varphi \rmd\vartheta \label{g11a}
\end{equation} 
which is to be evaluated  over the unit sphere in y-space where all eigenvalues are positive\footnote{One might think that this can easily be done with a computer algebra program. In our experience these programs nead a lot of assistance which is indicated by the following intermediate results. They are nevertheless of enormous value in confirming the results.}, 
i.e. the topmost triangle in figure 1. With
\[\det({\bf H})=r^3\Phi_3(\vartheta,\varphi)=r^3\left(\frac{\sqrt5\cos \vartheta} {6\sqrt6}(5-8\sin^2\vartheta)-
\frac{\sin^3\vartheta\cos3\varphi}{3\sqrt{6}}\right),\]
and the indefinite integral
\begin{eqnarray}
I  =
\int\Phi_3(\vartheta,\varphi)\Delta_3(\vartheta,\varphi)\sin\vartheta\,\rmd\vartheta =
I_1(\vartheta,\varphi)+I_2(\vartheta,\varphi), \nonumber \\
I_1(\vartheta,\varphi) = \frac{\sqrt{15}\sin 3\varphi}9\left(\frac{\sin^5\vartheta}4 -\frac{2\sin^7\vartheta}7\right)  \nonumber \\
I_2(\vartheta,\varphi) =  \frac{\sqrt3\sin 6\varphi\cos\vartheta}6
\left(\frac8{105}+\frac{4\sin^2\vartheta}{105}+\frac{\sin^4\vartheta}{35}
 +\frac{\sin^6\vartheta}{42}\right)
\label{g12}
\end{eqnarray} 
we obtain for the integral $I_{\rm mi}$ over the topmost spherical triangle
\begin{equation}
II_{\rm mi} = \int_0^{\pi/3}\left(I(\vartheta_1(\varphi),\varphi)-I(0,\varphi)\right)\,\rmd\varphi
=\int_0^{\pi/3}\left(I(\vartheta_1(\varphi),\varphi)\right)\,\rmd\varphi\label{g13} 
\end{equation} 
as the integral at $\vartheta=0$ vanishes obviously and where $\vartheta_1(\varphi)$ describes the great circle $E_1=0$, i.e. from \eref{g8}
\[ \sin\vartheta_1=\frac{\sqrt5}{\sqrt{5+4\cos^2\varphi}},\qquad
\cos\vartheta_1=\frac{2\cos\varphi}{\sqrt{5+4\cos^2\varphi}}. \]
Now the indefinite integrals $II_1(\varphi)$ and $II_2(\varphi)$ of $I_1$ and $I_2$  over $E_1=0$ can again be determined easily, and one obtains
\begin{eqnarray}
II_1(\phi) & = &
\frac{\sqrt{3}\cos(\varphi)}{189\sqrt{5+4\cos^2(\varphi)}^5}
\Big( -\frac{15}4 +8\cos^2(\varphi)-\frac{356\cos^4(\varphi)}{25}\Big)
\nonumber \\ II_2(\phi) & = &
-\frac{\sqrt{3}\cos^3(\varphi)}{9\sqrt{5+4\cos^2(\varphi)}^5} \Big(1
-\frac{48\cos^2(\varphi)}{25}- \\ & & \frac{816\cos^4(\varphi)}{875}+
\frac{3328\cos^6(\varphi)}{2625} +\frac{2048\cos^8(\varphi)}{2625}\Big).
\nonumber \end{eqnarray}
With these indefinite integrals we get for $II=II_1+II_2$
\begin{equation}
II(\frac{\pi}3)=-\frac{\sqrt2}{210},\quad
II(0)=-\frac{29\sqrt3}{3780},\quad
II_{\rm mi}=\frac1{210}\Big(\frac{29}{18}\sqrt3-\sqrt2\Big) \label{g14a}
\end{equation}
 and then also $n^{(2)}_{\rm mi}$ from \eref{g11a} where we have used the
 nomalization \eref{g11}
\[
n^{(2)}_{\rm mi}=\frac1{16\sqrt{\pi}\beta^{3/2}}\Big(\frac{29}{18}\sqrt3-\sqrt2\Big) \]
and these gives finally
\begin{equation}
n_{\rm mi} = \frac1{16\pi^2}\Big(\frac{\alpha}{\beta}\Big)^{3/2}\Big(\frac{29}{18}\sqrt3-\sqrt2\Big) 
=0.00872 \Big(\frac{\alpha}{\beta}\Big)^{3/2}
\end{equation}
for the spatial density of the maxima.

To determine the probability of the saddle points, we have to integrate over
the spherical triangle between the great circles $E_1=0$, $E_2=0$ and
$E_2=E_3$. The first one is given by $\vartheta=\vartheta_1(\varphi)$ and let
the second be given by $\vartheta=\vartheta_2(\varphi)$. From \eref{g8} we
obtain
\[ \sin\vartheta_2=\frac{\sqrt5}{\sqrt{5+4\cos^2(\varphi+\pi/3)}},\qquad
\cos\vartheta_2=-\frac{2\cos(\varphi+\pi/3)}{\sqrt{5+4\cos^2(\varphi+\pi/3)}}.\]
Then the integral over this triangle $II_{sa}$ can be written as
\begin{equation}
II_{\rm sa}=\int_0^{\pi/3}\left(I(\vartheta_1,\varphi)-I(\vartheta_2,\varphi)
\right)\,\rmd\varphi
=II_{\rm mi}-\int_0^{\pi/3}\left(I(\vartheta_2,\varphi)\right)\,\rmd\varphi
\label{g15} \end{equation}
where we have taken into account, that the integrand is negative between
$E_1=0$ and $E_2=0$. Now $\varphi=\tilde{\varphi}+2\pi/3$ gives
$\vartheta_2(\varphi)=\vartheta_1(\tilde{\varphi})$ and
$I(\vartheta_2(\varphi),\varphi) =
I(\vartheta_1(\tilde{\varphi}) ,\tilde{\varphi})$ and therefore
\[\int_0^{\pi/3}I(\vartheta_2,\varphi)\,\rmd\varphi =
\int_{-2\pi/3}^{-\pi/3}I(\vartheta_1(\tilde{\varphi}),
\tilde{\varphi})\,\rmd\tilde{\varphi} =
-\int_{\pi/3}^{2\pi/3}I(\vartheta_1,
\tilde{\varphi})\,\rmd\tilde{\varphi} = 2 II(\frac{\pi}3).  \]
We then obtain from \eref{g15} and \eref{g14a}
\[II_{\rm sa}=\frac1{210}\Big(\frac{29}{18}\sqrt3+\sqrt2\Big) \]
and for the ratio of the density of minima to that of critical points
\[\frac{II_{mi}}{2II_{mi}+2II_{sa}}=\frac{1}{4}-\frac{3\sqrt{6}}{58}=0.1233. \]

\section{Conclusion}

The distribution of critical points of a two-dimensional function is severely restricted by topological considerations. There the density of extremal points and saddle points agree, i.e. there is an equidistriburion of signs of the eigenvalues of the Hessian.
For three-dimensional functions these topological restrictions no longer apply and this ratio depends on the type of the function. For a Gaussian homogeneous and isotropic random function slghtly less than $1/8$ of all critical points are minima. A comparison with results from computational fluid dynamics seems interesting.

\ack
We are very grateful to H. Schanz and T. Schick for very helpful discussions. Furthermore we express our thanks to the anonymous referees of an earlier version of this manuscript for pointing out to us the important work done on the Hessians of homogeneous isotropic random functions.

\appendix

\section*{Appendix: Example functions}

A simple function with many critical points is given by
\begin{displaymath}
\cos\pi x_1 +\cos\pi x_2+\cos\pi x_3.
\end{displaymath}
Its critical points are the lattice points $(x_1,x_2,x_3)=(n_1,n_2,n_3)$ with integer $n_1$, $n_2$, $n_3$. Maxima occur when the integers are all even, minima when they are all odd. All other critical points are saddle points. Obviously all sign combinations are equidistributed. Under eight critical points one finds one maximum one minimum and 6 saddle points. We now indicate how functions with arbitrary ratios not smaller than one of densities of saddle points to extrema can be found. Actually one can conclude from Theorem 3.4 in the Morse theory chapter of\cite{l5a} that this restriction on the critical points is a necessary one. There are no functions having a density of extrema larger than the density of saddle points.

The purpose of this appendix is to describe functions with various ratios of saddle point density to extrema density. As a first step, we describe that it is under certain conditions possible to combine two functions to a new function which has exactly the critical points of both combined. To be specific, let $F({\bi x})$ and  $\Phi({\bi x})$ be two smooth functions with $F({\bi x}_0)=0$ and $\Phi(\infty)=0$ and furthermore
\[ |\nabla F({\bi x})|\leq M|{\bi x}-{\bi x}_0| \qquad  {\rm and} \qquad
|\nabla \Phi({\bi x})| \leq \frac{M}{{|\bi x}|^2}, \]
then with some nonvanishing vector $\bi V$ the functions 
\[f=a+{\bi V\cdot ({\bi x}-{\bi x}_0)}+F({\bi x}) \quad {\rm and}\quad \phi={\bi V\cdot x}+ \Phi({\bi x})\]
are such that $f$ has no critical points in a neighbourhood of ${\bi x}_0$ and $\phi$ has only a finite number of critical points. Let $\alpha(r)$ be a smooth non-increasing function which is one for $r\leq 1$ and vanishes for $r\geq 2$. Then one can define with some positive $\epsilon$
\begin{eqnarray*}
g & = & \alpha\Big(\frac{{\bi x}-{\bi x}_0}{\epsilon}\Big)\bigg(a+\epsilon^2\phi\Big(\frac{{\bi x}-{\bi x}_0}{\epsilon^2}\Big)\bigg)
+\Big(1-\alpha\Big(\frac{{\bi x}-{\bi x}_0}{\epsilon}\Big)\Big)f(x) \\
& = & a+V\cdot({\bi x}-{\bi x}_0)+\alpha\Big(\frac{{\bi x}-{\bi x}_0}{\epsilon}\Big) \bigg(\epsilon^2\Phi\Big(\frac{{\bi x}-{\bi x}_0}{\epsilon^2}\Big)-F({\bi x})\bigg)+F({\bi x}).
\end{eqnarray*} 
It is obvious that $g$ has for sufficiently small $\epsilon$ in $|{\bi x}-{\bi x}_0|\leq \epsilon$ just the critical points of $\phi$ and for $|{\bi x}-{\bi x}_0|\geq 2\epsilon$ just those of $f$. With standard estimates one can verify, that for small enough $\epsilon$ there are no critical points in the transition region $\epsilon \leq |{\bi x}-{\bi x}_0| \leq 2\epsilon$.

Now we describe functions having an arbitrary non-zero even number of saddle points and just two extrema. These functions can then be inserted in arbitrary numbers in some given smooth function and one can use them to generate functions having in any given domain an arbitrary rational ratio of saddle points to extrema above one. Then it is also possible to build functions with a similar density ratio.
These functions are necessarily three-dimensional. We nevertheless begin with a function of two variables $r,z$, namely
\[ f(r,z)=z+a\left(\frac1{1+b(r^2+(z+1)^2)}-\frac1{1+b(r^2+(z-1)^2)}\right) \]
with positive $a$ and $b$. Later on we will think of $r$ and $z$ as cylindrical coordinates in a three-dimensional space. The discussion of this function is elementary, so we just describe the results.  For small values of $a$ one has only a small perturbation of the linear function $z$ without any critical points.
For somewhat larger $a$ there are four critical points, one maximum, one minimum and two saddle points, all situated on the $z$-axis $r=0$. For even larger values of $a$, exactly for
\[4a>\left(b+\frac1b\right)^2 \]
the extrema remain on the $z$-axis while the saddle points move on the $r$-axis to $r=\pm r_s$ with some positive $r_s$ and this is the case which is of interest to us.  In three-dimensional space with cylindrical coordinates this function has however a non-generic critical circle $r=r_s$, so we would like to replace $a$ by $a\cos n \vartheta$, $\vartheta$ being the angle in cylindrical polar coordinates. This function is not even continuous at $r=0$, therefore we modify $a$ to some function $a(r^2)$. It is easily checked that the number and type of singularities is not changed if $a(r^2)$ is a monotonically decreasing positive function of $r$, which we assume. So, we have with
\[ f(r,z,\vartheta)=z+\frac{4bz(a_0(r^2)+a_1(r^2)cos(n\vartheta))}
{{(1+b(r^2+(z+1)^2))}{(1+b(r^2+(z-1)^2))}} \]
a smooth function which has a maximum and a minimum, both on the $z$-axis and $2n$ saddle points in the $z=0$ plane, provided $a_0(r^2)+a_1(r^2)cos(n\vartheta)$ is for all $\vartheta$ a monotonically decreasing positive function with
\[4a_0(0))>\left(b+\frac1b\right)^2\]
and $a_1(r^2)/r^n$ remains bounded at $r=0$.


\section*{References}

 \begin{thebibliography}{99}
\bibitem{l0} Wang L and Peters N 2006 The length-scale distribution function of the distance between extremal points in passive scalar turbulence.{\it J. Fluid Mech.} {\bf 554} 457--77
\bibitem{l00}  Loewen S, Ahlborn B and Filuk A B 1986 Statistics of surface flow structures on decaying grid turbulence.{\it Phys. Fluids} {\bf 29} 238--97
\bibitem{l01} Sreenivasan K, Prabhu A and Narasimha R 1983 Zero-crossings in turbulent signals. {\it J. Fluid Mech.} {\bf 127} 251--272.
\bibitem{l9} Moffatt H K 2001 The topology of scalar fields in 2d and 3d turbulence  {\it IUTAM Symp. on Geometry and Statistics of Turbulence} (ed. T. Kambe et al.)  Kluwer pp 13--22.
\bibitem{N1} Longuet-Higgins M S 1960 Reflection and Refraction at a Random Moving Surface. II. Number of Specular Points in a Gaussian Surface. {\it J. Opt. Soc. Am.} {\bf 50} 845-850.
\bibitem{N2} Halperin B L, Lax M 1966 Impurity Band Tails in the High-Density Limit I. Minimum Counting Methods. {\it Phys. Rev.} {\bf 148} 722-740.
\bibitem{N3} Rice S O, 1954 Mathematical Analysis of Random Noise in {\it Selected Papers on Noise and Stochastic Processes} edited by N. Wax. Dover
\bibitem{l1}  Mehta M L 2004 {\it Random Matrices} Elsevier
\bibitem{N4} Fyodorov Y V  2004 Complexity of Random Energy Landscapes, Glass Transition and Absolute Value of the Spectral Determinant of Random Matrices. {\it Phys. Rev. Lett.} {\bf 92} Art. No. 240601 and Erratum: {\it ibid.} {\bf 93} Art. No. 149901.
\bibitem{N4a} Fyodorov Y V  2005 Counting Stationary Points of Random Landscapes as a Random Matrix Problem {\it Acta Physica Polonica B} {\bf 36} 2699-2707.
\bibitem{N5} Bray A J and Dean D S 2007 The statistics of critical points of gaussian fields on large-dimensional spaces {\it Phys. Rev Lett.} {\bf 98} Art. No 150201.
\bibitem{N6} Fyodorov Y V, Sommers H-J and Williams I 2007 The density of stationary points in a high-dimensional random energy landscape and the onset of glassy behaviour {\it JETP Letters} {\bf 85} 261-266.
\bibitem{l4}  Davies H T 1962 {\it Introduction to Nonlinear Differential and Integral Equations} Dover
\bibitem{l5} Deimling K 1974 {\it Nichtlineare Gleichungen und Abbildungsgrade} Springer 
\bibitem{l5a} Hirsch M W 1976 {\it Differential Topology} Springer
\bibitem{l2}  Flanders H 1989 {\it Differential Forms with Applications to the Physical Sciences} Dover
\bibitem{l3} Mehta M L and Normand J-M  2001 Moments of the characteristic polynomial in the three ensembles of random matrices {\it J. Phys. A} {\bf 34} 4627-4639.
\end{thebibliography}
\end{document}